\documentclass[showpacs,showkeys,amsmath,amssymb,preprint,nofootinbib]{revtex4}
\usepackage{graphicx}
\usepackage{epsfig}
\usepackage{bm}
\usepackage{float}
\usepackage{fontenc}
\usepackage{graphicx}
\usepackage{bm}
\usepackage{amsmath}
\usepackage{xcolor}
\def\beq{\begin{eqnarray}}

\def\eeq{\end{eqnarray}}

\begin{document}

\title[PWH]{Modeling holographic dark energy with particle and future horizons}

\author{Miguel Cruz$^{a,}$}
\email{miguelcruz02@uv.mx}

\author{Samuel Lepe$^{b,}$}
\email{samuel.lepe@pucv.cl}

\affiliation{$^a$Facultad de F\'\i sica, Universidad Veracruzana 91000, Xalapa, Veracruz, M\'exico\\
$^b$Instituto de F\'\i sica, Facultad de Ciencias, Pontificia Universidad Cat\'olica de Valpara\'\i so, Av. Brasil 2950, Valpara\'\i so, Chile}

\date{\today}

\begin{abstract}
In this work we explore some cosmological properties coming from the particle and future horizons when are considered as candidates to model the dark energy sector within a holographic context in a flat Friedmann-Lemaitre-Robertson-Walker universe; we focus on some thermodynamics characteristics of the resulting dark energy scenario. Within the interacting scheme for cosmological fluids we obtain that in the dark sector, the dark energy fluid will always have negative entropy production and additionally, the positivity for the entropy and temperature can not be guaranteed simultaneously; this result holds for both horizons. However, this last issue can be solved if chemical potential is introduced in the thermodynamics description. For the non interacting approach, we obtain similar results as those of the single fluid description for the entropy behavior. We also find that the model admits a genuine big rip singularity when the dark energy density is sketched by the future horizon, in consequence the resulting parameter state can cross to the phantom regime. For the particle horizon case the cosmological fluid can emulate ordinary matter and the coincidence parameter has a decreasing behavior when the future horizon is elected.    
\end{abstract}


\maketitle
\section{Introduction}
\label{sec:intro}
If we consider that general relativity is the correct description of gravity at cosmological scales, it is well known that the observed expansion of the universe can be obtained from any homogenous and isotropic cosmological model, moreover, recent observations revealed that such expansion presents a late time accelerated phase \cite{riess}. The scale factor of a homogeneous and isotropic universe will expand with accelerated rate always that the pressure and density of the cosmological fluid obey the following relation, $p < -(1/3)\rho$. However, the acceleration could arise from an extra ingredient commonly termed as {\it dark energy} and possess the characteristic of having a negative pressure in order to have a repulsive gravitational effect. The simplest explanation for this accelerated expansion is provided by the cosmological constant but until now it has not been possible to provide a clear justification for the very small value of its magnitude. This and another kind of difficulties have led over the years to consider possible extensions or modifications of general relativity, from including a scalar field (see for instance Ref. \cite{copeland}) to increasing the number of dimensions of spacetime, see the Chapter 27 of Ref. \cite{pdg} for an interesting review on this subject. A complete compilation of proposals for dark energy models can be found in \cite{comp}, see also Ref. \cite{dynamical}, where the dynamical systems approach was used to study several dark energy models in standard cosmology and modified gravity.\\

An interesting approach for the dark energy problem is given by the holographic principle, which establishes that in quantum field theory the ultraviolet cut-off is related to the infrared cut-off, and this is imposed by the formation of a black hole \cite{cohen}. A complete review for holographic dark energy models can be found in \cite{holo1}. If $\rho$ is the density caused by the ultraviolet cut-off, then the total energy in a region of size $L$ must be of the order of the mass of a black hole of the same size, therefore such density must be of the form, $\rho \sim L^{-2}$, in agreement with the entropy-area relationship. On the other hand, if quantum corrections are considered in the holographic description, it is possible to obtain a generalization for the energy density and can be written as, $\rho \sim L^{2\delta -4}$, being $\delta$ a non-additivity parameter. In this case the entropy and the area of the black hole are related as $S \sim A^{\delta}$. Recent results show that this form of the density can be consistent with current observational data and exhibits a quintessence behavior \cite{npb1, npb2}. In general, the characteristic length $L^{-1}$ can be taken simply as the Hubble scale since the resulting density is comparable with the present day dark energy value \cite{holo1, hubble1, miao, mot1, mot2, cruz}. However, the Hubble scale is not the unique option to consider for the size, $L$, see the Refs. \cite{odintsov, odintsovde} where the particle and future horizons are considered as generalized holographic dark energy models for a specific $f(R)$ gravity model and for the unification between the inflationary stage and late time epoch in a scalar field cosmological model. In Ref. \cite{lee}, a holographic dark energy model based on the future horizon is explored in order to find a possible relation between the quantum entanglement and the dark energy. Within the holographic scheme a possible explanation for the discrepancy of the cosmological constant value is given in Ref. \cite{value} and such approach only invokes the symmetries of spacetime. On the other hand, in Ref. \cite{epjc} with the use of the apparent horizon, a holographic dark energy model is constructed for a curved universe, resulting that such approximation is adequate to describe the late time evolution of the universe. Also in Ref. \cite{newlepe} the holographic approach for a closed universe is investigated. These kind of holographic scenarios could result relevant to study since recent results show that a closed universe can provide a satisfactory physical explanation for the presence of an enhanced lensing amplitude in the cosmic background power spectra in the 2018 Planck legacy \cite{2019}.\\

In this paper we explore two different possibilities for the characteristic length, $L$, to be considered in the expression of the holographic dark energy density in order to establish which of the two options is a better candidate to describe this sector of the universe. As we will see later, the future horizon presents more similarities with the dark energy behavior in several aspects, being one of the most interesting that in this case the model admits a genuine big rip singularity (phantom scenario). It is interesting to have as final fate of the universe this scenario since is not ruled out by the observational data \cite{caldwell, des}. As shown in Ref. \cite{plhde}, the holographic dark energy models provide a good fit to the Planck data, besides the constrained value for the $c$-term appearing in the standard holographic formula for the energy density favors the phantom scenario. A more recent study revealed that within several holographic dark energy models, the future horizon case is the most favored by the observational data and exhibits an appropriate behavior at perturbative level since the growth of linear matter perturbations is also in agreement with observations \cite{holobs}. It is important to point out that if we consider the future horizon as dark energy model, we have an alleviation for the cosmological coincidence problem and on the other hand, the dark energy density becomes dominant over other possible components of the universe at late times. As we will see, for the particle horizon case the parameter state, $\omega$, can take positive values, therefore this characteristic length is less viable to describe the dark energy content of the universe. It is worthy to mention that some of the aforementioned results can be found in Ref. \cite{holo1}. Then, our aim in this work is to study some of the consequences at thermodynamics level for both possibilities of the characteristic length considered as dark energy candidates. As we will see below, when the interaction between the dark matter and dark energy is allowed, despite the temperature for the dark energy sector has a positive value, the entropy production is always negative and this leads to thermodynamics inconsistencies; this result is independent of the election for the characteristic length. However, for the phantom regime (allowed in the future horizon) we observe that the positivity problem of temperature and entropy can be solved with the inclusion of chemical potential, as discussed in other works \cite{termo3, lima1, saridakis, cruzlepeodintsov}, but again the entropy has a decreasing behavior; this contradicts the second law, which seems to be obeyed by the universe at cosmic scales as shown in Ref. \cite{manu}. In the non interacting scenario the obtained results are similar with those obtained in standard cosmology (single fluid description).\\ 

The organization of this work is the following: In Section \ref{sec:HDE} we provide some general aspects of the holographic dark energy for a spatially flat FLRW universe. We write some well-known results in the holographic scheme for the dark energy density and discuss the future singularity admitted by the model when the future horizon is considered. At the end of the section we briefly discuss the behavior of the dark energy density when the model approaches the far future. The Section \ref{sec:thermo} is devoted to the thermodynamics results for the holographic model, we adopt the standard thermodynamics point of view. The computed temperature for the particle horizon and future horizon remains positive but in the future horizon case has a divergent behavior in the far future. In Section \ref{sec:nonint} we consider an universe with dark energy and dark matter content with no interaction between them. From this description it is possible to see that the coincidence parameter has a decreasing behavior at late times, which is in agreement with observations, as expected, the adiabaticity condition is obtained for non interacting fluids. In Section \ref{sec:int} we consider the interacting scheme between dark energy and dark matter, the Section \ref{sec:entropy} is devoted to discuss the behavior of the entropy for dark energy and dark matter in the interacting scheme. In Section \ref{sec:chemical} we implement the introduction of chemical potential to solve the negativity problem of the entropy in the dark energy sector. In Section \ref{sec:final} we write the conclusions of our work.
     
\section{Holographic dark energy}
\label{sec:HDE}
In the framework of the spatially flat FLRW geometry, we define the Hubble rate as $H := \dot{a}(t)/a(t)$, where $a$ is the scale factor and the dot represents a derivative with respect to time. By means of the Friedmann constraint, we can write $3H^{2} = \rho$, being $\rho$ the energy density of the cosmological fluid. Since we are interested in a general holographic description, we will assume the conventional formula for the Hubble rate as 
\begin{equation}
H = \frac{c}{L},
\label{eq:hubble}
\end{equation} 
where $c$ is a positive constant for an expanding universe and $L$ is the cosmological length scale. For this radius we will focus on two possibilities, the particle and the future horizons denoted by $L_{p}$ and $L_{f}$, respectively, which are given by the following expressions
\begin{eqnarray}
L_{p}(t) &=& a(t)\int^{t}_{0}\frac{dt'}{a(t')}, \label{eq:particle}\\
L_{f}(t) &=& a(t)\int^{\infty}_{t}\frac{dt'}{a(t')}.\label{eq:future}
\end{eqnarray}  
If we consider the Eqs. (\ref{eq:hubble})-(\ref{eq:future}), we can obtain \cite{odintsov, odintsovde}
\begin{equation}
\frac{d}{dt}\left(\frac{c}{a(t)H} \right) = \pm \frac{1}{a(t)},
\end{equation}
where the sign $+(-)$ corresponds to the particle (future) horizon. From the previous expression we can solve for the Hubble rate
\begin{equation}
\dot{H} + \left(1\pm \frac{1}{c} \right)H^{2} = 0,
\end{equation}
whose solution is given by
\begin{equation}
H(t) = \frac{H(t_{0})}{1+\left(1\pm 1/c \right)H(t_{0})(t-t_{0})},
\label{eq:hubblehor}
\end{equation}
being $t_{0}$ some initial time. Using the above equation we can write the following expression for the scale factor
\begin{equation}
a(t) = a(t_{0})\left[1+\left(1\pm \frac{1}{c} \right)H(t_{0})(t-t_{0}) \right]^{1/(1\pm 1/c)}.
\label{eq:scalefactor}
\end{equation}
On the other hand, if we consider a barotropic equation of state, $p = \omega \rho$, where $p$ and $\rho$ are the pressure of the fluid and its density respectively, together with the Friedmann equations and the Eq. (\ref{eq:hubblehor}), the parameter state takes the form
\begin{equation}
\omega^{\pm} = -1 - \frac{2\dot{H}}{3H^{2}} = - 1 + \frac{2}{3}\left\lbrace 1 \pm \frac{1}{c}\right\rbrace.
\label{eq:omega}
\end{equation}
In general, $c$ is a constant given in the interval $0 < c < 1$ \cite{miao}, therefore we can have the following possibilities: $1 + 1/c > 1$ and $1 - 1/c < 0$, it is important to point out that in each case the parameter state given in Eq. (\ref{eq:omega}) will represent an ordinary matter behavior or a phantom fluid, respectively. Besides, for the choice $c = 1$ in the above equation results, $\omega^{-} = -1$, which represents a cosmological constant evolution, this result is also obtained in Ref. \cite{miao}. However, note that only under the election of the future horizon, the model resembles a cosmological constant evolution. A more general description for the holographic dark energy can be found in Ref. \cite{radicella}, where was suggested that $c$ is in fact a slowly varying function in the interval $0 < c(t) < 1$. Therefore, as the universe expands we have that the $c(t)$-term variates within the interval $(0,1)$, in such case we can observe from Eq. (\ref{eq:omega}) that the parameter state, $\omega^{-}$, at some stage of the cosmic evolution will behave as a phantom fluid or could describe a cosmological constant like evolution depending on the value taken by the $c$-term, i.e., the phantom behavior is only a transient stage if we consider the value $c$ as a function. A transient phantom scenario can be also found in the DGP brane model \cite{dgp} or when other effects are considered in the standard cosmology such as particle production in an universe permeated by a phantom fluid \cite{transient2}, for example. An interesting feature of a holographic description given by a variable $c$-term is its consistency with observations at background and perturbative levels, this results can be found in Ref. \cite{variable}.\\ 

If we consider the future horizon case in the expression (\ref{eq:hubblehor}) we have
\begin{equation}
H(t) = \frac{H(t_{0})}{1+\left(1/c - 1\right)H(t_{0})(t_{0}-t)},
\end{equation}
by defining $t_{0} = t_{s} - 1/[(1/c -1)H(t_{0})]$ in the previous expression one gets
\begin{equation}
H(t) = \frac{1}{1/c - 1}(t_{s}-t)^{-1},
\end{equation}
which represents a genuine big rip singularity for $t = t_{s}$ according to the classification given in Ref. \cite{odintsov1}, besides $t_{s} = t_{0} + 1/[(1/c -1)H(t_{0})] > t_{0}$, by means of the Friedmann equations $\rho \sim H^{2}$ and $p \sim 2\dot{H}+3H^{2}$, therefore as $t \rightarrow t_{s}$ we have $\rho \rightarrow \infty$ and $p \rightarrow \infty$ simultaneously. $t_{s}$ represents a finite time in the future at which the singularity will take place. The generalities of a big rip singularity within the holographic context was studied in Ref. \cite{elizalde}. In Ref. \cite{cruz} can be found that within the framework of holographic dark energy a singular behavior can be induced for the Hubble rate when the spatial curvature is included. However, the singularity obtained is not a genuine big rip but instead a Type III future singularity.\\ 

By considering the standard expression for the redshift, $1+z = a(t_{0})/a(t)$, from the scale factor given in Eq. (\ref{eq:scalefactor}) we can write
\begin{equation}
(1+z)^{-\alpha_{\pm}} = 1+\alpha_{\pm}H(t_{0})(t-t_{0}),
\end{equation}    
where $\alpha_{\pm}$ is a constant defined as $\alpha_{\pm}:=1\pm 1/c$. By replacing the previous equation in the expression (\ref{eq:hubblehor}), we obtain the Hubble rate as a function of the redshift as follows
\begin{equation}
H_{\pm}(z) = H(0)(1+z)^{\alpha_{\pm}},
\label{eq:ratered}
\end{equation}
where $H(0)$ is a constant, if we consider $0 < c < 1$, as the model approaches the far future $(z=-1)$, we have $H_{+}(z\rightarrow -1)\rightarrow 0$ and $H_{-}(z\rightarrow -1)\rightarrow \infty$. This represents a main difference between this model and the $\Lambda$CDM model, where the Hubble rate remains bounded for the future evolution.\\ 

In order to describe the cosmological fluid we will consider the particle and future horizons as holographic cut-off, therefore the conventional formula for the dark energy density becomes
\begin{equation}
\rho = 3c^{2}L^{-2}_{p,f},
\end{equation} 
where $p$ and $f$ denotes the particle (future) horizon, respectively. The previous expression can be written as a function of the redshift as follows
\begin{equation}
\rho^{\pm}(z) = \rho^{\pm}(0)(1+z)^{2\alpha_{\pm}}, 
\label{eq:redde}
\end{equation}
where the Eqs. (\ref{eq:hubble}) and (\ref{eq:ratered}) were considered together with the Friedmann constraint and we have defined the constant $\rho^{\pm}(0):=3c^{2}H^{2}(0)$. It is important to point out an interesting feature of the density expression given in Eq. (\ref{eq:redde}), if we consider the future horizon case we can see that the exponent will be negative, therefore as the model evolves to the future this density increases, $\rho^{-}(z\rightarrow -1)\rightarrow \infty$, in other words, the dark energy density becomes dominant over other matter components. This characteristic for the dark energy is identified in the well known cosmological coincidence problem and it is corroborated by observational data \cite{velten}. On the other hand, for the particle horizon case, as the model evolves to the future the density tends to zero.  

\section{Thermodynamics}
\label{sec:thermo}
In standard cosmology, for a perfect fluid we have the following temperature evolution equation 
\begin{equation}
\frac{\dot{T}}{T} = -3H\left(\frac{\partial p}{\partial \rho}\right)_{n}.
\label{eq:evolution}
\end{equation}
The previous equation is valid always that the Gibbs integrability condition holds together with the number ($n$) and energy conservation. However, as can be seen, the constantcy of the temperature is no longer available, i.e., the global equilibrium condition, $\dot{T} = 0$, is not satisfied \cite{maartens}. If we consider a barotropic equation of state in the temperature evolution we can write
\begin{equation}
\int d \ln T = 3 \int \omega(z) \frac{dz}{1+z} \ \Rightarrow \ T(z) = T(0)\exp \left(3 \int \omega(z) \frac{dz}{1+z} \right),
\label{eq:tempe}
\end{equation}
where $T(0)$ is an integration constant. If in the previous expression we consider the parameter state given in Eq. (\ref{eq:omega}), we obtain for the temperature
\begin{equation}
T^{\pm}(z) = T(0)(1+z)^{3\omega^{\pm}},
\label{eq:temperature}
\end{equation}
from the last equation we can obtain the following condition, $T^{+}(z \rightarrow -1) \rightarrow 0$, and for the future horizon we have a divergent behavior as we approach the far future. Given that $\rho^{-}$ has an increasing behavior together with $T^{-}$, we will assume that when the dark energy is described by the future horizon, the thermodynamics description is consistent.\\ 

As can be seen from the previous results, the temperature associated to the holographic dark energy is always positive independently of the choice of the particle or future horizon. Within the context of standard thermodynamics, the problem of the entropy and temperature for the phantom regime has been widely studied, since the positivity of both quantities can not be guaranteed simultaneously unless a negative chemical potential is introduced by hand \cite{termo3, lima1, saridakis}. This problem was solved recently in Ref. \cite{cruzlepeodintsov} by the introduction of dissipative effects in the framework of irreversible thermodynamics.\\

\subsection{Non interacting cosmological fluids}
\label{sec:nonint}
From now on, we will denote by $\rho_{1}$ the dark energy density and $\rho_{2}$ will describe a dark matter density, we will also assume the corresponding parameter state for the dark matter as $\omega_{2} = 0$. For an universe composed by dark energy and dark matter, the Friedmann constraint reads
\begin{equation}
3H^{2}(z) = \rho_{1}^{\pm}(0)(1+z)^{2\alpha_{\pm}} + \rho_{2}(0)(1+z)^{3},
\end{equation}
where the dark energy density is described by the holographic expression given in Eq. (\ref{eq:redde}) and $\rho_{2}(0)$ is an appropriate constant. The corresponding continuity equations for the densities are given as follows
\begin{align}
& (\rho^{\pm}_{1})' - 3 \left(\frac{1+\omega^{\pm}_{1}}{1+z} \right)\rho^{\pm}_{1} = 0,\label{eq:noninteracting1}\\
& \rho'_{2} - \left(\frac{3}{1+z} \right)\rho_{2} = 0.
\label{eq:noninteracting2}
\end{align} 
where the prime stands for a derivative with respect to $z$, additionally we have
\begin{equation}
r^{\pm}(z) = r^{\pm}(0)(1+z)^{3-2\alpha_{\pm}},
\label{eq:coincidence}
\end{equation}
being $r^{\pm}(z)$ the coincidence parameter, which is defined as the ratio between the densities for dark matter and dark energy, $r^{\pm}(z) = \rho_{2}(z)/\rho^{\pm}_{1}(z)$ and we have defined the constant $r^{\pm}(0) := \rho_{2}(0)/\rho_{1}^{\pm}(0)$. In order to be in agreement with observational data, the coincidence parameter must decrease as the universe expands \cite{velten, wang}, note that as we approach to the far future ($z=-1$), the coincidence parameter given in Eq. (\ref{eq:coincidence}) can have an increasing (decreasing) behavior, which only depends on the election of the particle (future) horizon\footnote{For the particle horizon case, when $c < 2$, we have a singular behavior for the coincidence parameter when $z=-1$.}. This is consistent with the result obtained for the holographic dark energy density in the previous section. The coincidence parameter (\ref{eq:coincidence}) can be written in terms of the scale factor, one gets in each case
\begin{align}
& r^{+} = r^{+}(0)\left(\frac{a}{a_{0}} \right)^{\frac{2}{c}-1},\\
& r^{-} = r^{-}(0)\left(\frac{a_{0}}{a} \right)^{1+\frac{2}{c}},
\end{align}   
since $0 < c < 1$, we have the limits $r^{+}(a\rightarrow \infty) \rightarrow \infty$ and $r^{-}(a\rightarrow \infty) \rightarrow 0$. From the previous results we can observe that the coincidence parameter problem can be solved in the context of holographic dark energy under the election of the future horizon. It is worthy to mention that this differs from the results obtained in Refs. \cite{coincidence1, coincidence2}, where an interacting scheme between holographic dark energy and a pressureless dark matter is considered, in such cases the ratio between densities remains constant.\\

On the other hand, in terms of the future horizon the Gibbs equation reads  
\begin{equation}
T^{-}dS^{-} = d(\rho^{-}_{1} V) + p^{-}_{1}dV = d[(\rho^{-}_{1}+p^{-}_{1})V]-Vdp^{-}_{1} =  Vd\rho^{-}_{1} + \rho^{-}_{1}(1+\omega^{-}_{1})dV,
\label{eq:gibbs}
\end{equation}
where $V$ is the Hubble volume given by $V(a)=V(a(t_{0}))(a(t)/a(t_{0}))^{3}$, therefore $dV/V = 3(a(t)/a(t_{0}))^{-1}d(a(t)/a(t_{0})) = 3Hdt$ and also we have considered a barotropic equation of state, yielding
\begin{equation}
T^{-}\frac{dS^{-}}{dt} = V\frac{d\rho^{-}_{1}}{dt} + \rho^{-}_{1} (1+\omega^{-}_{1})\frac{dV}{dt},
\end{equation} 
and by means of the continuity equation for the density $\rho_{1}$ one gets
\begin{equation}
T^{-}\frac{dS^{-}}{dt} = 0,
\label{eq:adiabatic}
\end{equation} 
therefore we can see that, $S = \mbox{constant}$, and this is independent of the election of the particle (future) horizon for the dark energy density $\rho_{1}$, this feature is generally obtained for non interacting systems and it is consistent with the cosmological constant evolution. However, as pointed out in Ref. \cite{victor}, the cosmological constant scenario lacks of physical consistency at thermodynamics level, therefore in the next section we consider the interacting scheme. It is important to point out that the conservation equation for the entropy obtained in (\ref{eq:adiabatic}) or simply, $\dot{S} = 0$, means that there is not entropy flux, therefore $S$ is constant for each particle of the fluid and this is consequence of the conservation of the particle number $n$, i.e., $\dot{n}+3Hn = 0$ \cite{maartens}. To end this section, it is worthy to mention that the Gibbs equation written in (\ref{eq:gibbs}) is given in terms of the internal energy, $\rho V$. Despite this is the most common expression for such energy, it leads to inconsistencies if one demands the simultaneous fulfillment of the Friedmann and thermodynamic pressure equations in an expanding universe. However, an alternative expression for the energy that alleviates the aforementioned problem is the Komar energy with the following form, $(\rho + 3p)V$, with a generalized continuity equation $\dot{\rho}+3(\dot{p}+3Hp)+3H(\rho + p)=0$. This form of the energy has the virtue of considering the contribution of the pressure in the fluid, leading to a more consistent thermodynamics scenario to describe the dark sector. This thermodynamic consistency can be also guaranteed in some holographic approaches \cite{komar1}. In order to strength our thermodynamics description we could consider the Komar energy instead the internal energy, but as stated in Ref. \cite{komar2}, the internal energy is only valid in a flat universe, which corresponds to the case considered in this work. 

\subsection{Interacting cosmological fluids}
\label{sec:int}
For two interacting cosmological fluids, the continuity equations for both densities given in the expressions (\ref{eq:noninteracting1}) and (\ref{eq:noninteracting2}) must be written as follows
\begin{align}
& (\rho^{\pm}_{1})' - 3 \left(\frac{1+\omega^{\pm}_{1}}{1+z} \right)\rho^{\pm}_{1} = \frac{Q}{H(z)(1+z)}, \label{eq:q1}\\
& \rho'_{2} - \left(\frac{3}{1+z} \right)\rho_{2} = -\frac{Q}{H(z)(1+z)}, \label{eq:q2}
\end{align} 
where the $Q$-terms determine the interaction between the dark energy and dark matter and can be expressed as functions of the redshift. In general, the $Q$-terms are given by differents Ansatze and can be elected by convenience. The cosmological imprints of this interaction can be probed with the use of observational data and everything seems to indicate that $Q > 0$ \cite{wang, Q, Q2}, this means that there exists an energy flow from dark energy fluid to dark matter one \cite{victor}. For $Q<0$ we have transference of energy from dark matter to dark energy. See also Ref. \cite{int1}, where was found that the interacting scheme for holographic dark energy evolves in agreement with observational data and exhibits stability, and Ref. \cite{int2}, where the inclusion of spatial curvature in the interacting scheme leads to an expanding and accelerating universe.\\     

If we consider the Gibbs equation given in (\ref{eq:gibbs}) for the equations (\ref{eq:q1}) and (\ref{eq:q2}) separately, we can write
\begin{equation}
\frac{Q}{H(z)(1+z)} = \frac{T_{1}}{V}\frac{dS_{1}}{dz} = - \frac{T_{2}}{V}\frac{dS_{2}}{dz}.
\label{eq:interaction}
\end{equation}
On the other hand, the equation (\ref{eq:q1}) can be written in equivalent form as follows
\begin{equation}
(\rho^{\pm}_{1})' - 3 \left(\frac{1+\omega^{\pm}_{1,eff}}{1+z} \right)\rho^{\pm}_{1} = 0,
\end{equation}
where we have defined, $\omega^{\pm}_{1,eff}:= \omega^{\pm}_{1}+Q/3H(z)\rho^{\pm}_{1}$, if we insert this effective parameter state in Eq. (\ref{eq:tempe}), we can write
\begin{equation}
T(z) = T(0)(1+z)^{3\omega^{\pm}_{1}}\exp \left( \int \frac{Q(z)}{\rho^{\pm}_{1}(z)H(z)(1+z)}dz\right),
\label{eq:tempdark}
\end{equation}   
then, the positivity of the temperature for the dark energy sector does not depend on the direction of the energy transference between this and the dark matter one. Besides, as obtained in the non interacting case, this result is independent of the election of the future or particle horizon.

\subsection{Entropy behavior}
\label{sec:entropy}
In this section we will provide a general description of the entropy behavior for the {\it dark cosmological sector} \cite{zim} obtained from Eq. (\ref{eq:interaction}).\\
$\bullet$ {\bf Dark matter sector:}\\
For this case we have that $\omega_{2} = 0$, therefore from the Euler relation 
\begin{equation}
TS = (1+\omega)\rho V,
\label{eq:euler}
\end{equation}
we can establish $T_{2}S_{2} = \rho_{2}V > 0$, then from this last result we have that $S_{2} > 0$ and $T_{2} > 0$ by means of the Eq. (\ref{eq:interaction}), considering the change, $z \rightarrow t$, one gets
\begin{equation}
\frac{dS_{2}}{dt} > 0,
\end{equation}
therefore for the dark matter sector we have a growth in the entropy.\\
$\bullet$ {\bf Dark energy sector:}\\
In this case the Euler relation (\ref{eq:euler}) reads, $T^{\pm}_{1}S^{\pm}_{1} = (1+\omega^{\pm}_{1})\rho^{\pm}_{1}V$, and given that the parameter state, $\omega^{-}_{1}$, can take values in the phantom regime, therefore $T^{-}_{1}S^{-}_{1} < 0 $, according to Eq. (\ref{eq:interaction}) we have two possibilities: $T^{-}_{1} < 0, dS^{-}_{1}/dt > 0$ and $T^{-}_{1} > 0, dS^{-}_{1}/dt < 0$, but since we are in the interacting scheme, if the entropy of the dark matter sector grows, the corresponding entropy to the dark energy must decrease, then the first possibility is discarded if we consider the obtained previous result, therefore we are left with the condition
\begin{equation}
\frac{dS^{-}_{1}}{dt} < 0,
\end{equation}
with positive temperature, then for the dark energy sector the entropy decreases while for the dark matter sector it increases. Note that the product $T^{-}_{1}S^{-}_{1}$ is negative, thus, based on the previous arguments we must have, $S^{-}_{1} < 0$, and this is the aforementioned positivity problem of the entropy and temperature for the phantom universe. 

\subsection{Including chemical potential in the interacting scheme}
\label{sec:chemical}
As found previously, from Eq. (\ref{eq:tempdark}) we can observe that the temperature associated to the dark energy sector will be always positive, therefore we now follow the line of reasoning of Refs. \cite{termo3, lima1, saridakis, cruzlepeodintsov}, which requires the introduction of chemical potential in order to solve the problem of negative entropy. Given that we are considering the interacting scheme we have entropy production, as can be seen in Eq. (\ref{eq:interaction}). In Ref. \cite{maartens} it is established that when no other effects are considered in the fluid; the particle production must contribute to the generation of entropy, i.e., we have non conservation of the particle number, $n$. Once that the number of particles of a system it is included in the thermodynamics description, we must take into account the role of the chemical potential. If we have a particle production rate given as, $\nu$, the non conservation condition can be written as follows
\begin{equation}
\frac{\dot{n}}{n}+3H = \nu = \frac{\dot{N}}{N},
\end{equation}  
where $nV = N$, being $V$ the volume containing $N$-particles, and the chemical potential, $\mu$, it is introduced by the expression \cite{callen}
\begin{equation}
T^{\pm}_{1}dS^{\pm}_{1} = d(\rho^{\pm}_{1} V) + p^{\pm}_{1}dV - \mu dN.
\label{eq:gibbs2}
\end{equation}
If we consider a barotropic equation of state and perform the time derivative of the previous expression, one gets
\begin{equation}
\dot{\rho^{\pm}_{1}}+3H\rho^{\pm}_{1}(1+\omega^{\pm}_{1}) = \frac{1}{V}\left(\mu \dot{N}+ T^{\pm}_{1}\frac{dS^{\pm}_{1}}{dt} \right),  
\end{equation}
where we used the relation $\dot{V}/V = 3H$. If we consider the change of variable, $z \rightarrow t$, together with Eqs. (\ref{eq:q1}) and (\ref{eq:interaction}), we can have an explicit expression for the chemical potential given as
\begin{equation}
\mu(t) = - \frac{1}{\nu N}\left(QV-T^{\pm}_{1}\frac{dS^{\pm}_{1}}{dt} \right) = - \frac{2Q(t)}{\nu n(t)},
\label{eq:chemical}
\end{equation}
Note that the chemical potential only depends on the interaction term $Q$ and the particle number, $n$. In dependence of the sign of the $Q$-term and the production ($\nu > 0$) or annihilation of particles ($\nu < 0$), the chemical potential remains negative or turns positive. On the other hand, from Eq. (\ref{eq:chemical}) we can observe that the chemical potential is a function of time (or the redshift); this differs from the standard description of thermodynamics when the microscopic motion of particles forming a body becomes relativistic. In such description the temperature remains constant together with the chemical potential, which is given as $\mu = \mu_{0} + mc^{2}$, being $mc^{2}$ the rest energy of a particle of the body. However, in order to obtain this behavior for the temperature and chemical potential, it must be assumed that the system has reached the thermal equilibrium (even in presence of external fields) \cite{landau}. Then, given that our thermodynamics description does not obey the global equilibrium condition, $\dot{T} = 0$, as can be seen in Eq. (\ref{eq:evolution}), we can say that certain differences with respect to the standard scheme are expected to appear.\\ 
With the introduction of chemical potential and a barotropic equation of state, we can write the Euler relation in the following form
\begin{equation}
T^{\pm}_{1}S^{\pm}_{1} = (1+\omega^{\pm})\rho^{\pm}_{1}V - \mu N, 
\end{equation}
and as previously found, for $\omega^{-}_{1}$ we can have a phantom behavior. However, given that the chemical potential is negative, if the following condition for the phantom regime is fulfilled
\begin{equation}
\mu > -(1+\omega^{-}_{1})\frac{\rho^{-}_{1}}{n},
\end{equation}
or equivalently, $Q > -\nu (1+\omega^{-}_{1})\rho^{-}_{1}$, then the product $T^{-}_{1}S^{-}_{1}$ will be positive leading to $T^{-}_{1} > 0$ simultaneously with $S^{-}_{1} > 0$. As commented previously, the chemical potential seems to have an important role at cosmological level in the phantom regime. A recent work shows that if the dark sector is decoupled from the Standard Model fields, one way to assign it a temperature under certain considerations is given by the introduction of chemical potential, in consequence the dark matter could be warm or cold \cite{chemical}. It is worthy to mention that despite the fact of having positive entropy and positive temperature for the phantom regime, given that we are considering an interacting scheme, the entropy must have a decreasing behavior. Therefore, these thermodynamics issues could be solved by the inclusion of some other effects in the fluid or by extending the standard thermodynamics scheme.\\ 
As final comment for this section, note that from the Eq. (\ref{eq:interaction}) we can have
\begin{equation}
\frac{d}{dt}\left(S_{1}+S_{2} \right) = \frac{VQ}{T_{2}}\left(1-\frac{T_{2}}{T_{1}}\right), \ \Rightarrow \ \frac{d}{dt}\left(S_{1}+S_{2} \right) \gtrless 0, \ \ \mbox{only \ if} \ \ T_{2} \lessgtr T_{1}.  
\end{equation}
As pointed out in Ref. \cite{maartens}, the symmetries of the FLRW spacetime allow only scalar dissipation, i.e., absence of energy flux due to heat flow, therefore the total entropy of the system fulfills the second law of thermodynamics always that $T_{2} < T_{1}$. However, until now the determination of both temperatures values remains as a challenge for modern cosmology. 
  
\section{Final remarks}
\label{sec:final}
In this work we consider a holographic description to model the dark energy in a spatially flat FLRW universe. This description is based on a comparison between the particle horizon and future horizon. Under this construction it was obtained a general solution for the Hubble rate, the information of each horizon can be identified in this general solution by the election of its corresponding sign, $+$, for the particle horizon and, $-$, for the future horizon. According to the election of the sign and the value of the constant $c$, which enters in the conventional holographic formula, the Hubble rate has several limit cases as the model approaches the far future, this is an important difference between this model and the $\Lambda$CDM model. Specifically, when the future horizon case is considered, the model admits a genuine big rip singularity.\\

By considering a barotropic equation of state we can extract the general form of the parameter state, $\omega$, which also has the information of each horizon through the election of the appropriate sign, in a consistent way, when the future horizon is elected, the parameter state can take values in the phantom regime, $\omega < - 1$, additionally, for the specific value $c = 1$, the model resembles a cosmological constant evolution; within the holographic scheme of variable $c$-term, the phantom regime could be a transient stage of cosmic evolution.\\

In order to have a clear visualization of the behavior of all quantities, we introduced the cosmological redshift in each of them. In this way, from the Friedmann equations we can see that the density of the dark energy content has an increasing behavior as the model approaches the far future when the future horizon is considered. This is the desired behavior for the dark energy since the observations also indicates this growth and it is known as the cosmological coincidence problem. On the other hand, for the particle horizon case, the density has a diluting behavior.\\

If we consider that the universe besides the dark energy also contains a dark matter fluid, we can provide a description when these fluids can interact or not. By adopting the second case we can construct the corresponding coincidence parameter and again, we have a decreasing behavior when the future horizon is considered and in complementary way, when the coincidence parameter is written in terms of the scale factor, we can see that as the universe expands the coincidence parameter tends to zero. From a thermodynamics point of view, the temperature for the dark energy fluid when it is described by the future horizon tends to increase as the cosmic evolution takes place and could also become singular at the far future, $z\rightarrow -1$. On the other hand, for the particle horizon case we can have a phase of cooling down as we approach to the far future. For both cases the entropy takes a constant value. This resembles the adiabatic expansion in a single fluid description. When we consider an interacting scheme between the holographic dark energy and dark matter we can observe that in such case the entropy production for the dark energy sector will be negative for both horizons despite the positivity of its temperature. The positivity of this temperature it is independent of the direction in which flows the energy in the dark sector. Given that the holographic dark energy exhibits a phantom regime when the future horizon is considered, by means of the Euler relation we found that in such case the entropy is negative. In order to solve the negativity problem of the entropy in the dark energy sector, we introduced chemical potential in the thermodynamics description, we found some differences with the standard relativistic thermodynamics since the resulting chemical potential is a function of time or the redshift and only depends on the interaction $Q$-term and the particle number $n$. The chemical potential can be positive or negative, being the latter the most interesting case, the change in the sign of the chemical potential it is mediated by the $Q$-term and by the production or annihilation of particles. It is worthy to mention that a negative chemical potential leads to a positive entropy for the dark energy sector. However, a decreasing behavior for this entropy is still obtained and this contradicts the second law. This suggests that we must improve the description of the fluid in order to have a consistent thermodynamics description for the interacting holographic dark energy. For instance, in the context of irreversible thermodynamics and considering dissipative cosmological effects as it was done in Ref. \cite{cruzlepeodintsov}, the resulting scenario is free of thermodynamics inconsistencies.\\ 

In general grounds, the particle horizon is not a good candidate to model the dark energy content of the universe. One might think that in such case it could be a good alternative to model dark matter, if we consider a cosmological model with dark energy ($\rho^{-}_{1}(z)$) and dark matter ($\rho^{+}_{1}(z)$), then the Friedmann constraint will be given by
\begin{equation}
3H^{2} = \rho_{1}^{-}(0)(1+z)^{2\alpha_{-}} +  \rho_{1}^{+}(0)(1+z)^{2\alpha_{+}},
\end{equation} 
by definition the coincidence parameter is $r(z) = \rho^{+}_{1}(z)/\rho^{-}_{1}(z) \sim (1+z)^{4/c}$, therefore $r(z\rightarrow -1)\rightarrow 0$, which is in agreement with observational data. Additionally, for the particle horizon we have, $\omega^{+}_{1} = -1 + 2(1+1/c)/3$, if we consider the value $c=1$ we obtain, $\omega^{+}_{1} = 1/3$, and for $c=1/2$ the fluid behaves as stiff matter. These values for the parameter state can not describe an accelerating universe.\\

Finally, it is important to mention that holographic models for dark energy are supported by the observational data and exhibit a consistent matter growth perturbations and stability \cite{holo1, holobs, variable, int1}.

\section*{Acknowledgments}
M.C. work has been supported by S.N.I. (CONACyT-M\'exico).

\end{document}